\documentclass[twocolumn]{aastex631}

\usepackage{graphicx}	
\newcommand{\ergs} {erg s$^{-1}$}
\newcommand{\Porb}{$P_{\rm orb}$}
\usepackage{amsmath}	
\usepackage{booktabs}  
\usepackage{natbib}
\usepackage{float}
\usepackage{hyperref}
\usepackage{tabularx}

\begin{document}

\title{On the Nature of the BeWD System Nova FQ\,Circinus}
\correspondingauthor{Shatakshi Chamoli}
\email{shatakshi.chamoli@iiap.res.in, shatakshichamoli@gmail.com}

\author[0009-0000-5909-293X]{Shatakshi Chamoli}
\affiliation{Indian Institute of Astrophysics, 2nd Block Koramangala, 560034, Bangalore, India}
\affiliation{Pondicherry University, R.V. Nagar, Kalapet, 605014 Puducherry, India}

\author[0009-0000-5909-293X]{Philip A. Charles}
\affiliation{School of Physics and Astronomy, University of Southampton, Southampton SO17 1BJ, UK}
\affiliation{Department of Physics, University of the Free State, 205 Nelson Mandela Drive, Bloemfontein, 9300, RSA}
\affiliation{Astrophysics, Department of Physics, University of Oxford, Keble Road, Oxford OX1 3RH, UK}

\author[0000-0001-7570-545X]{Judhajeet Basu}

\affiliation{Indian Institute of Astrophysics, 2nd Block Koramangala, 560034, Bangalore, India}

\author[0000-0003-3533-7183]{G.C. Anupama}
\affiliation{Indian Institute of Astrophysics, 2nd Block Koramangala, 560034, Bangalore, India}

\author[0000-0002-3927-5402]{Sudhanshu Barway}
\affiliation{Indian Institute of Astrophysics, 2nd Block Koramangala, 560034, Bangalore, India}

\author[0000-0001-7016-1692]{Przemek Mr\'oz}
\affiliation{Astronomical Observatory, University of Warsaw, Al. Ujazdowskie 4, 00-478 Warszawa, Poland}

\begin{abstract}

The classical nova FQ\,Cir \citep{Schaefer26} has recently been shown to have a Be star donor, and has been identified as a High Mass Cataclysmic Variable (HMCV), with an orbital period of $\sim 2.04\,d$ (twice the 1.02\,d periodogram peak). Using the extensive OGLE optical archive, together with existing observations from  ASAS-SN we instead find significant modulation between $\sim 50-60$~days and propose that the orbital period is likely closer to $\sim 62\,d$. We interpret the low amplitude 1.02\,d modulation as a non-radial pulsation of the Be star. We also reassess the HMCV classification based on properties of known Be star-white dwarf systems.
\end{abstract}


\section{Introduction}
\begin{figure*}
    \centering
    \includegraphics[scale=0.43]{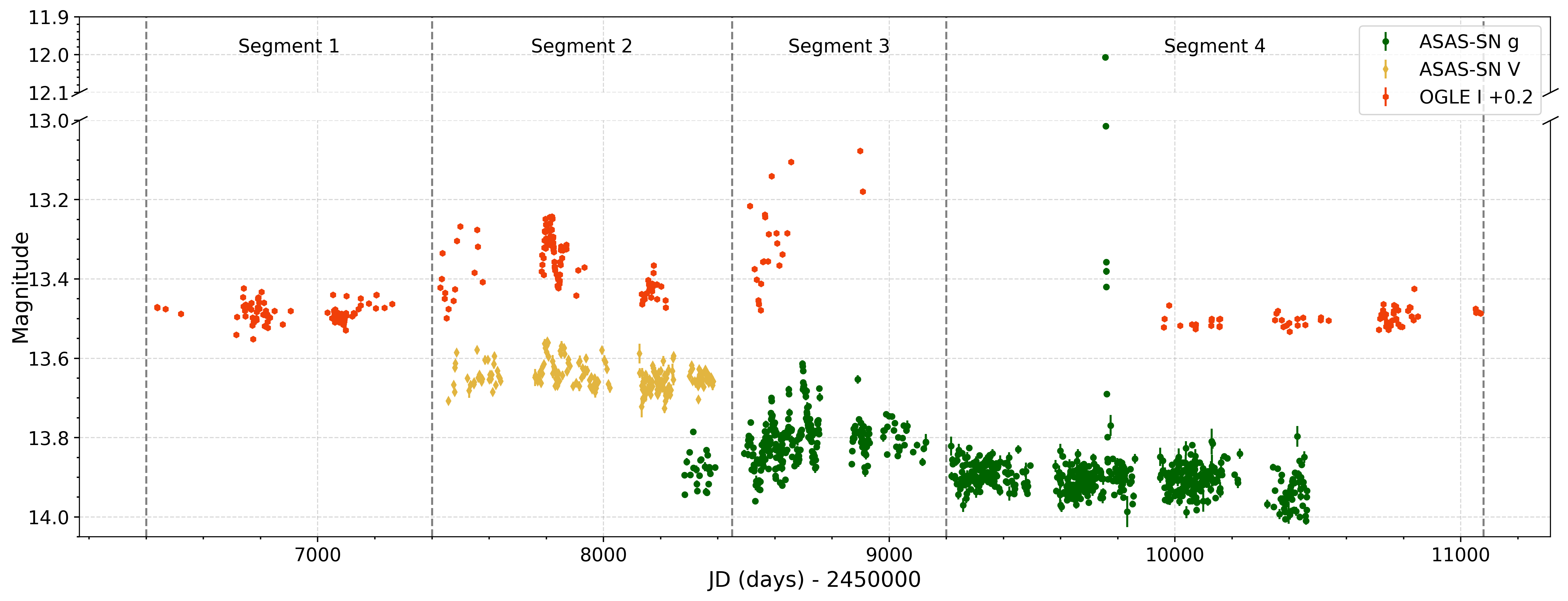}
    \caption{$\sim$12-year photometric light curves of FQ\,Cir obtained from ASAS-SN and OGLE (see section \ref{sec: data} for details). The 2022 nova outburst of FQ\,Cir is clearly visible at day $\sim$9800.}
    \label{fig:full_LC}
\end{figure*}

Both forms of novae (classical and recurrent) are members of the class of cataclysmic variables (CVs), which are short orbital period (typically a few hours to a few days) interacting binaries where a cool, usually main sequence, low-mass star \citep{2012ApJ...746...61D} transfers matter onto a white dwarf (WD) via an accretion disc \citep{1995cvs..book.....W}. Novae are characterized by thermonuclear outbursts observed as a dramatic, large change in brightness (6-14 mags, \citealt{2015ApJ...811...32P}), followed by a (usually) long-term (months to years) decline to quiescence. These outbursts are straightforward to distinguish from the lower amplitude ``dwarf novae'', which are a property of the CV's accretion disc changing from a cool, optically thin state to one that is hot and viscous (see e.g. \citealt{Lasota2001}).

During quiescence the optical emission is usually dominated by the secondary (mass-losing) star and the bright spot (where the mass-transfer stream from the secondary impacts the accretion disc), whereas in the UV there are contributions from the accretion disc and the WD. 

Orbital periods of known CVs are usually around a few hours, but can cover the range from $\sim5$ minutes to a few days (see \citealt{2026MNRAS.548ag514D} and references within), although longer period systems are now being revealed by long-term surveys (e.g. \citealt{Mroz2024}). There is also a class of accreting WD systems with (very) much longer orbital periods (months to years), called ``symbiotic novae'', where the mass-donor is an evolved red giant. These too show thermonuclear outbursts, and can have much longer durations than typical novae \citep{2025CoSka..55c..47M}.

Recently a new class of WD binaries have also been identified with massive donor stars, called Be/white dwarf systems (or BeWD), see Table \ref{tab:BeWDprops} for details and references. As a result of their binary evolution, classical Be stars are rapidly rotating, which gives rise to an outflowing, dense, wind-focussed disc at the equator (which is often described as an ``equatorial or decretion disc''). This is seen as characteristic Balmer emission features in their 
optical spectra (see \citealt{Porter2003}, \citealt{2013A&ARv..21...69R}). In BeWDs, the
proposed mode of accretion onto the WD is through mass transfer from the decretion disc. Most of the BeWDs have been discovered as the result of the transient ``supersoft'' (SSS) X-ray phase that is frequently seen during the decline of a nova outburst. Given the intrinsically high optical brightness of
the Be star, the amplitude of the optical nova peak is greatly reduced ($\leq3$ mags), likely explaining why they have been so difficult to find. That we now know of a dozen or so BeWDs is almost entirely through their transient SSS components, and we shall give more details on these systems in section 3.


Furthermore, nova outbursts have been spectroscopically confirmed in two BeWDs; a rapidly recurrent nova system in M31 \citep{shafter-2024, 2025ApJ...991..174C}, and a Galactic Be system, FQ\,Cir  \citep{Schaefer26}, that is the subject of this paper. 

 Be stars show photometric and spectroscopic variability over a broad range of timescales, arising from both the central star and the disc \citep{2022AJ....163..226L}. Analogous to CVs, the decretion disc shows disc outbursts on varying, long timescales. However, these discs are transient and dissipate and re-form over very extended timescales. The Be star itself also contributes to the observed variability on shorter timescales through both stellar rotation and pulsation activity \citep{Rivinius2013}. In a binary, additional variability may arise from orbital motion and interactions between the companion and the decretion disc (see section 4.4 in \citealt{2017AJ....153..252L}).

Unlike Be/neutron star systems (BeXRB), the formation of BeWDs does not involve a supernova kick, and the orbits remain circular \citep{Raguzova}. Since Be stars show photometric modulations on various timescales, searching for the orbital period becomes more challenging. From photometric data alone the orbital period can only be known for certain if the system is observed at high inclination and is eclipsing. 

Discovered through its fast nova outburst in June 2022 \citep{2022CBET.5138....1P}, the BeWD system FQ\,Cir was proposed as the prototype ``High Mass Cataclysmic Variable'' by \citealt{Schaefer26} (hereafter S26). What marked it out as unusual was the 2.04 day, 0.028\,mag modulation from the TESS light curve which S26 proposed to be the orbital period. Here we present an alternate interpretation for this modulation, identify a possible orbital period from Optical Gravitational Lensing Experiment \citep[OGLE,][]{2015AcA....65....1U} and  All-Sky Automated Survey for Supernovae \citep[ASAS-SN,][]{2023arXiv230403791H,2014ApJ...788...48S} datasets and also comment on the HMCV classification.

\begin{figure*}
    \centering
    \includegraphics[scale=0.6]{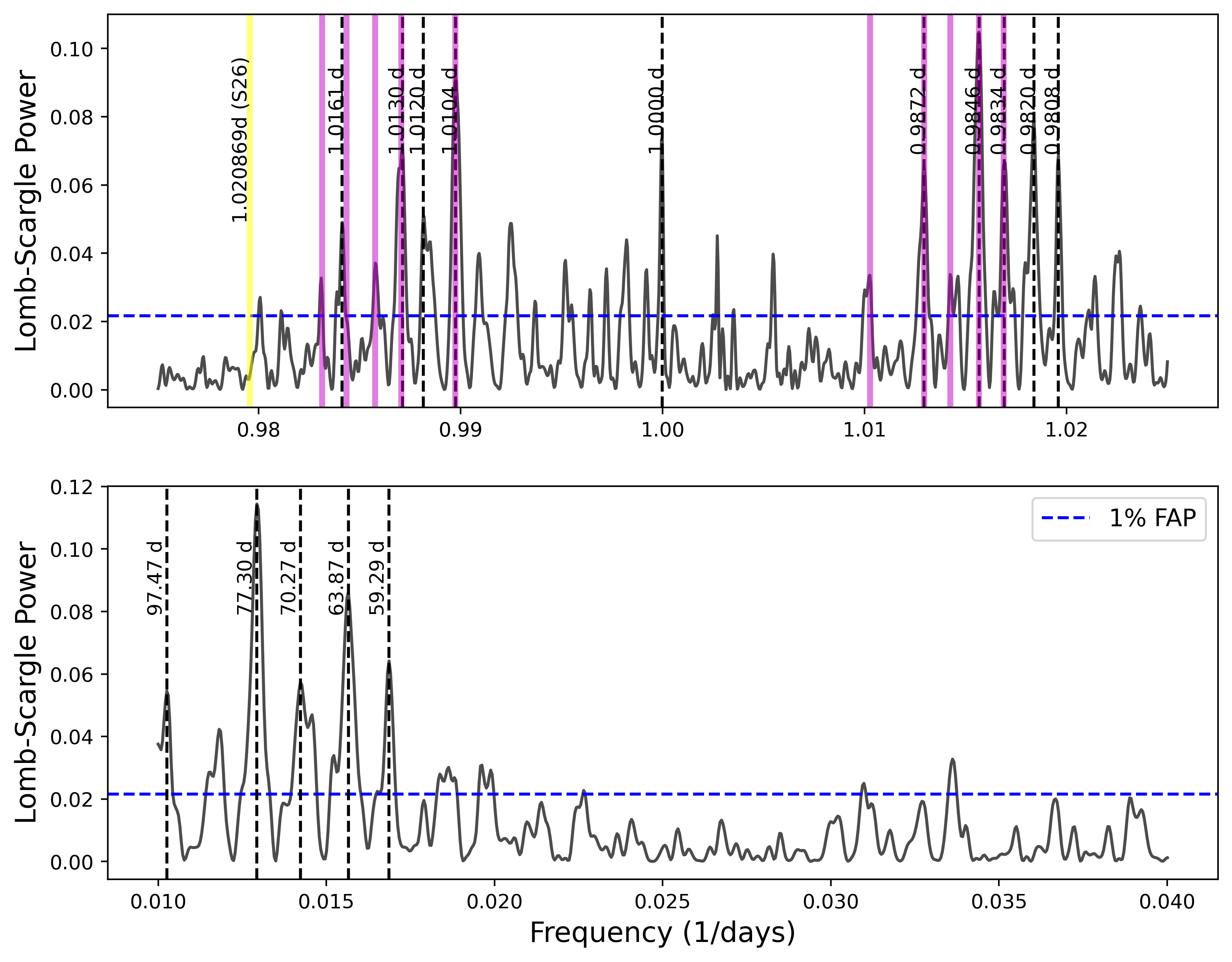}
    \caption{Lomb-Scargle periodogram for the combined, de-trended photometry of FQ\,Cir. Vertical black dashed lines indicate peaks well in excess of the 1\% FAP level (blue horizontal dashed line) and are labeled with the corresponding period. Magenta lines correspond to alias frequencies of those significant peaks with P$>$10\,d. The yellow line marks the TESS period found by S26.}
    \label{fig:LSP}
\end{figure*}

\begin{figure*}
    \centering
    \includegraphics[scale=0.43]{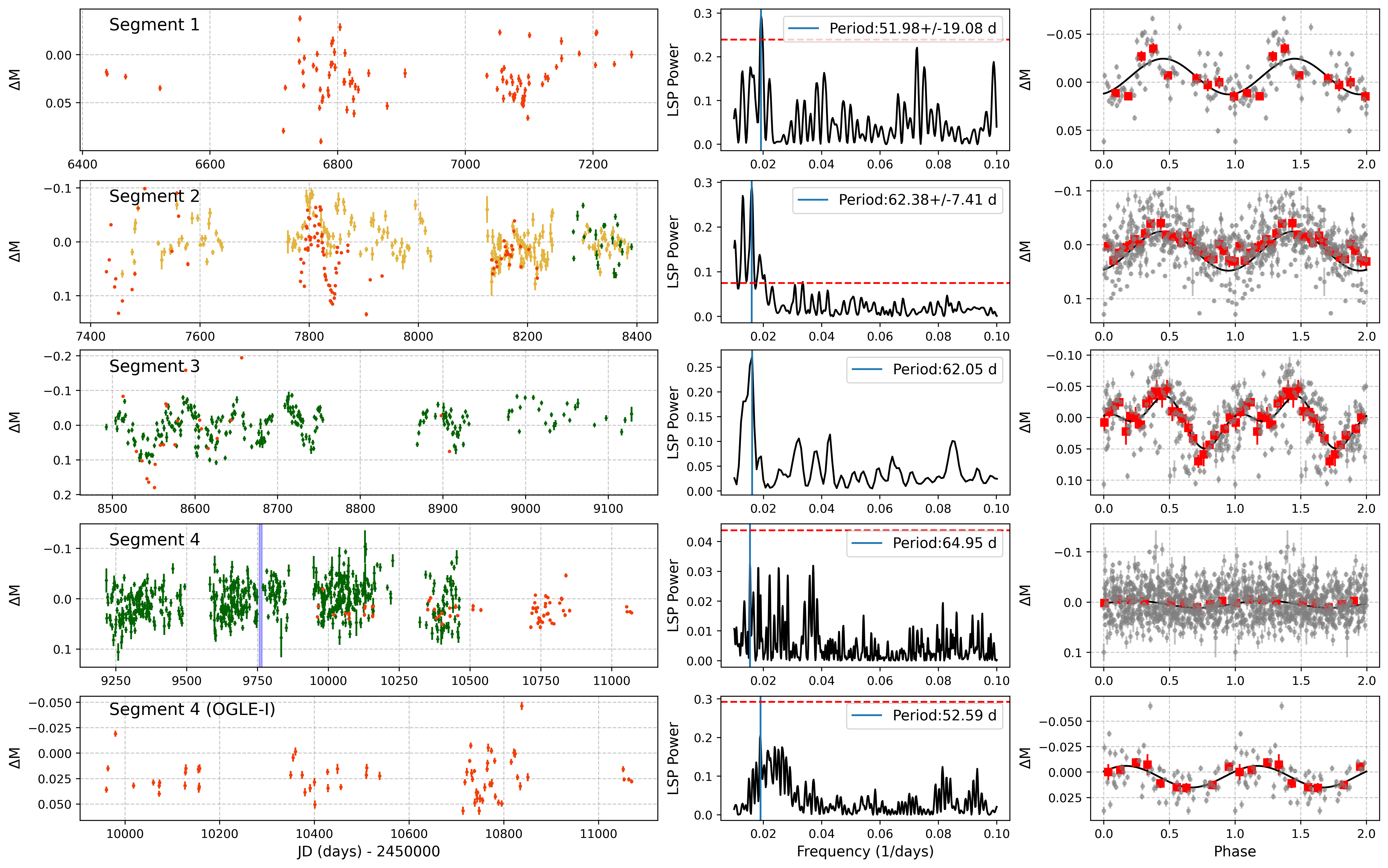}

    \caption{Segment-wise search for modulation in detrended ASAS-SN and OGLE light curve. Left: detrended light curves (using same color scheme as in Fig. \ref{fig:full_LC}). The blue line marks the duration of the nova outburst (data points within this interval were removed for the analysis). Middle: Lomb-Scargle periodogram; blue vertical line: peak power; dashed red line: 1\% FAP level. Right: Phase-folded light curve (on each periodogram peak period); red marker: binned data set; black line: Lomb-Scargle model.}
    \label{fig:segment LC}
\end{figure*}

\section{Data Analysis} \label{sec: data}

We use data from OGLE (I-band) and ASAS-SN (V- and g-band, from the ASAS-SN Sky Patrol Photometry Database) spanning 12.6 and 9.25 years respectively. The ASAS-SN dataset also covers the 2022 outburst. We divide the light curve into four segments based on the activity level (see Figure \ref{fig:full_LC}):
\begin{itemize}
    \item Segment 1 (day\,6400-7400): The light curve shows minimal activity with a maximum amplitude of 0.13 mag in the OGLE-I band. 
\item Segment 2 (day\,7400-8400): the OGLE-I shows multiple $\sim2$ months long outbursts, where one showed a maximum amplitude of 0.26 mags. 
\item Segment 3 (day\,8500-9200): the light curve reaches its peak in the whole dataset. We see a long term modulation superposed over the smaller variations in the ASAS-SN-g band and multiple $\sim2$ months long outbursts in the OGLE-I band. 
\item Segment 4 (day\,9200-11200): apart from the nova outburst, the light curve shows a similar variation to that seen in segment 1, the maximum amplitude of modulation is $\sim0.2$ mags in the ASAS-SN-g band, and $\sim0.12$ mags in OGLE-I. 
\end{itemize}

To facilitate a search for periodic signals, we combine the ASAS-SN and OGLE light curves to effectively increase the baseline. This is done by de-trending the light curves to remove long-term variations and applying appropriate color shifts. For the ASAS-SN-g dataset, we fit a weighted smoothing spline to the light curve and subtract it to remove the long-term trend while preserving the shorter-timescale variability. We also perform a 3-sigma clipping to remove the data corresponding to the nova outburst. A color shift is applied to the ASAS-SN-V-band to match the ASAS-SN-g-band median. For the OGLE dataset, clusters of dense observations were identified. Color shift for each cluster was calculated by comparing the mean magnitude of the cluster with the detrended ASAS-SN-g median. Now all the data is at the same median level, and \textgreater 10 day modulations can be recovered more effectively. 

We use the Lomb-Scargle Periodogram to search for modulations in the combined dataset. The top 5 most significant peaks above the $1\%$ false alarm probability level in the periodogram (see Figure \ref{fig:LSP}) correspond to 77.30, 64.05, 59.29, 70.27 and 97.47 days . We also search for modulations in the four individual segments of the combined light curve and separate filter datasets whenever the combined dataset fails to provide satisfactory fits (see Figure \ref{fig:segment LC}). In segments 1 and 2, a clear modulation of $\sim52$ days and $\sim62$ days, respectively, can be identified. In segment 3, the light curve exhibits an uneven double-peaked modulation. The best fit for this feature is obtained using the ASAS-SN-g data and increasing the number of fourier terms in the periodogram to two. In segment 4 no clear modulation is apparent in the combined dataset, which is mainly the ASAS-SN-g observations. We therefore examine the OGLE-I dataset separately and find a period of $\sim51.28$ days, similar to that obtained in segment 1. Although this period is not significant, likely due to a lack of data, the modulation can be clearly seen in the detrended light curve.

We also check for modulations around the 1 day period reported by S26 (see Figure \ref{fig:LSP}). However, it is difficult to recover such a  period in our dataset due to the 1-day sampling of both OGLE and ASAS-SN. Moreover the frequency space around the 1-day period is dominated by aliases.

\section{Discussion} \label{diss}

\begin{table*}
    \footnotesize
    \caption{Properties of BeWD Systems}
    \label{tab:BeWDprops}
    \begin{tabularx}{\textwidth}{l c c c c c c c l}
    \hline
    Source & Host & Companion & {$V$} & \Porb & Outburst Type  & log $L_X$ & WD Mass & Ref. \\
    & &  & (mag) & (d) & & (\ergs\ ) & ($M_\odot$) & \\
    \hline

    XMMU\,J052016.0$-$692505 & LMC & B0$-$3e & 14.80 & 510 or 1040 & SSS & 36.74 & 0.9--1.0 & [1] \\
    
    RX\,J0527.8$-$6954 & LMC & B5e V & 17.90 & $>0.875$ & Long SSS transient & 36.8$\pm$2 & 1.14--1.4 & [2] \\
    
    XMMU\,J010147.5$-$715550 & SMC & O7IIIe$-$B0Ie & 14.47 & 1264 & Long SSS transient &  34.48 & -- & [3],[4] \\
    
    MAXI\,J0158$-$744 & SMC & B1/2IIIe & 14.90 & $439_{-80}^{+114}$ & Short SSS outburst & $>$39 & $\sim$1.35 & [4]-[6] \\
    
    Swift\,J004427.3$-$734801 & SMC & O9V--B2III & 15.00 & 21.5 & Long SSS transient & $36.61^{+0.14}_{-0.33}$ & -- & [7] \\
    
    Swift\,J011511.0$-$725611 & SMC & O9IIIe & 14.36 & 17.402 & Long SSS transient & $36.52\pm0.04$  & $\sim$1.2 & [8] \\

    CXOU\,J005245.0$-$722844 $^\dagger$ & SMC & O9$-$B0e & 14.94 & 17.55 & Short SSS transient & $38.81^{+0.14}_{-0.09}$ & 1.23 & [9]-[10] \\
    
    EP250630b & LMC & Be star & 15.70 & $\sim13$ or $\sim40$ & Short SSS transient & $\sim 37$& -- & [11] \\
    
    {[HP99]}\,159 & LMC & He star& $\sim$16.0 & 2.327 & Persistent SSS & 36.8 & 1.2 & [12],[13] \\
    
    M31N\,2017$-$01e & M31 & O9.7 $-$ B0.6e & 20.50 & 14.3 & Recurrent nova & -- & $>1.3$ & [14],[15] \\
    
    \hline
    
    FQ\,Circini & MW & B1 V(n)(e) & 14.01 & 2.041738(?) & Classical nova & -- &1.25 & S26 \\
    
    \hline
    
    \end{tabularx}
    \footnotesize
    Notes: $^\dagger$ Also known as EP J005245.1$-$722843 \\
    References: [1] \cite{Kahabka-bewd1}, [2] \citet{Oliveira-bewd6}, [3] \cite{Sturm-bewd2}, [4] \citet{li-bewd3}, [5] \citet{mori-bewd3}, [6] \citet{2024RNAAS...8...32C}, [7] \citet{Coe-bewd5}, [8] \citet{Kennea-bewd4}, [9] \citet{cxou-bewd7}, [10] \citet{ep-bewd8},   [11] \citet{2025ATel17428....1M}, [12] \citet{Greiner23}, [13] \citet{Szegedi26}, [14] \citet{2025ApJ...991..174C}, [15] \citet{shafter-2024}

\end{table*}

\subsection{Origin of the 1.02 d modulation}

Be stars show variability at various timescales, amplitudes and duration \citep{2011MNRAS.413.1600R, 2017AJ....153..252L}. The lowest amplitude and shortest period are of the order of $\sim\,0.1-2$ days, which are attributed to stellar non-radial pulsations and rotation. Intermediate timescale (\textgreater 2\,d) variability is generally quasi-periodic, and is related to decretion disc outbursts, binary interaction etc (see Section 4.4 in  \citet{2017AJ....153..252L}). The long term periodicity (months to years) arises due to secular change in the decretion disc (formation or destruction) or due to superorbital modulations. 

Simulations of Be star discs show that tidal truncation does not prevent disc formation in the presence of a close companion, and that the extent of the disc decreases with the orbital separation/period, which should be easily observable. This suggests that the scarcity of short-period systems is not a selection effect but a consequence of binary evolution \citep{2016MNRAS.461.2616P}. Binary population synthesis shows that, for BeWDs with orbital periods less than ~10 days, tidal synchronization causes the Be star to spin down, leading to disc dissipation \citep{2014ApJ...796...37S}. Other physical mechanisms, such as unstable mass transfer leading to stellar merger, and stable mass transfer leading to expansion of the orbit, also do not favor the formation of short period systems.

Population synthesis models predict the orbital period of BeWD to be in the range of 20-600\,d \citep{zhue-bewd}. However, it is difficult to recover such a period as it overlaps with other intermediate timescale variability from the Be star disc. The shortest reported orbital period of a Be-binary is of MX\,Pup at 5.15\,d \citep{2002A&A...385..488C}, and for a Be+compact object is 12\,d for SAX\,J2103.5+4545 \citep{2007MNRAS.374.1108B}, a Be+NS system. At 2.04\,d FQ\,Cir would have the shortest orbital period among all known Be-binary systems. This very low amplitude, short-term variation can be seen in the TESS light curve of FQ\,Cir (see Fig 2 in S26), and similar features are also seen by TESS in other Be star light curves \citep{2022AJ....163..226L}. It is well known that pulsation is a ubiquitous phenomenon in Be stars, with periods in the range of hours to 2 days, and with very low amplitudes \citep{2003A&A...411..229R, 2017AJ....153..252L}, which are very unlikely as an orbital modulation. Hence, we propose that the true origin of the 1.02\,d period in FQ\,Cir is non-radial pulsations, which naturally accounts for the very low modulation amplitude.

 S26 also argue for the 1.02\,d period based on their spectra, which shows weak filling of the Balmer lines. This spectral feature does indicate that the extent of the disc at that time is small, and it can be reasoned that the disc would be truncated by the presence of a companion. However, all the S26 spectra were taken post the 2022 nova outburst, after which the light curve returned to its ``quiescent'' baseline level (see Figure \ref{fig:full_LC}). Given the low amplitude photometric variability during this time, implying that the disc activity is low, the spectra more likely indicate that the decretion disc has dissipated. Accordingly, we predict that the Balmer lines will show strong emission components once the significant optical variations return.

It is interesting to note that the pre- and post-nova brightness of the system is practically identical (see OGLE-I in segments 1 and 4, and ASAS-SN-g between MJD 9800 and 10000 in Figure \ref{fig:full_LC}), implying that the nova outburst had little or no impact on the disc. This is consistent with the fact that the WD in the system is massive and the mass of the ejected envelope is low \citep{Lechuga_2025}.

\subsection{A longer orbital period?}

More importantly, we note that between MJD 8540 and 8800 (see Figures \ref{fig:full_LC} and \ref{fig:segment LC}), the system is at its brightest and shows periodic modulation with uneven minima, similar to the ellipsoidal modulation seen in Roche-lobe filling binary systems. Given that the inclination of the system is not large, and that brightening can be associated with the growth of the disc, this would imply that the disc is at its maximum extent. In this configuration the disc material can be accreted by the WD in a pseudo-Roche-lobe overflow fashion. Moreover the system can exhibit ellipsoidal modulation due to the deformed shape of the decretion disc. We believe this is what is shown in the light curve. It is also interesting to note that the nova outburst occurs after these months-long disc outbursts. 

We estimate a period of 62.05\,d for this variability, during the brightest phase in the light curve. We also search for modulations in the full light curve and detect peaks between 50 and 100 days. As mentioned earlier in this section, estimating the orbital period of BeWD systems from photometric data is not straightforward. What we essentially observe is the effect of the companion on the disc. These interactions can be seen in the disc as ellipsoidal precession, the propagation of tidally locked density waves, and disc deformation, among others \citep{2017AJ....153..252L}. Given the plethora of physical mechanisms that give rise to these intermediate periods, the periodogram is very complex. Hence, we can only obtain a rough, indirect estimate of the orbital period from the light curve. Based on our analysis, we estimate it to be around 62.05\,d.

\subsection{Comparison with other BeWD Systems}

To effect a comparison of the properties of FQ\,Cir with the small group of BeWDs already known, we have collected them together in Table \ref{tab:BeWDprops}. All of these systems have hinted at the presence of a WD with a Be star, either through SSS transient outbursts (both long and short duration) and through nova outbursts.

Reported, or suggested, orbital periods for BeWDs span a wide range (see Table~\ref{tab:BeWDprops}). Systems such as Swift\,J004427.3-734801, Swift\,J0011511.0--725611, CXOU\,J005245.0-722844, M31N\,2017-01e and EP250630b have been proposed to have orbital periods between $\sim$10-20\,d, whereas XMMU\,J052016.0-692505, XMMU\,J010147.5-715550 and MAXI\,J0158-744 have shown \textgreater 500\,d modulations. These much longer periods quoted may not be orbital at all as the systems need to be in closer orbit to allow for accretion. Systems in orbits $\leq2$\,d like [HP99]159 are highly evolved, constantly accreting and are persistent SSS sources. As in the case of FQ\,Cir, we believe these orbital period estimates are not the true orbital period but only approximations. This is especially evident in the case of CXOU\,J005245.0-722844, which showed an apparent change in period between two observing cycles. This behaviour was interpreted as a consequence of precession in an expanded, warped/tilted decretion disc rather than any change in the companion's orbit \citep{cxou-bewd7}. Such effects are a well-known feature in certain types of cataclysmic variables (e.g. \citealt{Sun24}) and the resulting modulation is called a ``negative superhump''.  This takes place in the accretion disc around the WD, but the physical effect will apply equally well to the circumstellar disc around the Be star. More importantly, such superhumps always occur at periods a few percent shorter than $P_{orb}$, thereby providing a valuable indicator.

Furthermore, Be stars are also found in binary systems with non-compact objects. A growing number of Be + O/B subdwarf (BesdOB) systems have been identified, with orbital periods between ~28-129 days \citep{2023AJ....165..203W, 2022MNRAS.516.3366N}. These estimates come from spectroscopy or interferometry, but there is only one case, V658 Car, which is an eclipsing binary, where an accurate orbital period could be derived from the light curve \citep{2025ApJ...995...87D}. Detecting orbital modulations associated in systems with such low mass compact objects can be difficult as the effect of the subdwarf on the disc will be relatively weak.

Both BeWDs and CVs share similar photometric variability such as disc outbursts, flickering, disc precession, nova outbursts etc. Like CVs, it is now confirmed that the Be star can host magnetic WDs, as in the case of $\gamma$\,Cas \citep{2026A&A...707A.334N} and likely its analogs. However, the origin of this variability is very different in the two classes. In BeWDs, variability primarily originates due to changes in the decretion disc of the Be star, whereas in CVs it is driven by the accretion disc of the WD. In CVs the mass is transferred through Roche lobe overflow from the donor star via an accretion disc to the WD. In BeWDS it is the decretion disc that transfers mass to the accretion disc of the WD. This disc-to-disc transfer is unique to BeWDs and BeXRBs. Moreover, it is well-established that Be star decretion discs are transient. And unlike CVs, the donor (the Be star+disc) does not fill its Roche lobe. This is also the case with symbiotic novae, which are also considered distinct from CVs. It is clear that the BeWDs are a type of high-mass accreting WD system. But given the fundamental distinction of the discs in the two systems we strongly caution against classifying BeWDs as HMCVs.

\section{Conclusion}
We analyzed the range of modulations observed in the light curve of the nova-producing BeWD system, FQ\,Cir. We propose that the 1.02\,d modulation detected in  TESS data by S26 is likely the NRP period rather than orbital in origin. This interpretation is supported by studies showing that Be stars are pulsators, commonly with periodic modulations of $<$2 days. Moreover, Be binaries, with extended circumstellar discs, are not expected to form with extremely short orbital periods.

Our analysis of the long term OGLE and ASAS-SN light curves reveals months long periodic modulations superposed over even longer variations. In segment 3 of the light curve, where FQ\,Cir is at its brightest phase, suggesting that the circumstellar disc is at its maximum extent, we detect a modulation with uneven minima at a period of 62.05\,d. We propose that during this bright phase, the modulation is linked with its companion WD, allowing the orbital period to be inferred. As the WD accretes from the disc, the disc may become warped as it extends, potentially producing ellipsoidal modulation. We therefore suggest that the true orbital period is close to the 62.05\,d modulation seen during this bright phase, and we predict that the Balmer lines will show strong emission components once the significant optical variations return.

\section*{Acknowledgment}
We thank the anonymous reviewer for their comments and useful suggestions.
PAC thanks the Leverhulme Trust for the support of an Emeritus Fellowship. GCA thanks the Indian National Science Academy (INSA) for support under the INSA Senior Scientist Programme.

\facilities{OGLE, ASAS-SN}

\software{\texttt{Python v3.10.12}, 
         \texttt{astropy v6.1.7} \citep{astropy:2022},
         \texttt{scipy v1.15.1} \citep{2020SciPy-NMeth}
         \texttt{NumPy v1.26.4} \citep{harris2020array},
         \texttt{pandas v2.2.0} \citep{pandas},
         \texttt{matplotlib v3.9.3} \citep{Hunter:2007}
        }

\bibliography{reference}

\end{document}